\documentclass[12pt]{article} 
\usepackage{amssymb,amsmath,amsfonts} 
\usepackage[dvips]{graphicx}
\usepackage{epic,curves}
\begin{document} 
\begin{titlepage}
\title{Non-unitary observables
in the 2d critical Ising model} 

\author{Louis-Pierre Arguin\thanks{E-mail:
    arguinl@crm.umontreal.ca}, Yvan Saint-Aubin\thanks{E-mail: saint@crm.umontreal.ca}\\Centre de recherches math\'ematiques \\and\\ 
D\'ept. de math\'ematiques et de statistique\\Universit\'e de
    Montr\'eal\\C.P.\ 6128, succ. centre-ville\\Montr\'eal, Qu\'ebec\\Canada H3C 3J7}
 \maketitle

\end{titlepage}

\begin{abstract} 
We introduce three non-local observables for the two-dimensional
Ising model. At criticality, conformal field theory may be used
to obtain theoretical predictions for their behavior. These formulae
are explicit enough to show that their asymptotics are described by
highest weights $h_{pq}$ from the Kac table for $c=\frac12$ distinct from
those of the three unitary representations ($0, \frac1{16}$ 
and $\frac12$).
\end{abstract} 

 \section{Introduction}

It is widely agreed that the Ising model in two dimensions is
described at its critical temperature by a conformal field
theory at $c=\frac12$. The spectrum of the transfer matrix
is that of the operator $L_0\oplus \overline{L_0}$ and is described
by a linear combination of the squared norm of characters of the
three unitary representations at this value of the central
charge, namely those with highest weights $0$, $\frac1{16}$
and $\frac1{2}$. The exact linear combination depends on the
boundary conditions put on the geometry under consideration.

These highest weights belong to Kac table $h_{pq}=(((m+1)p-mq)^2-1)/
4m(m+1)$ at $c=1-6/m(m+1)$. For minimal models, the
relevant (unitary) weights $h_{pq}$ are labeled by the integers $p, q$
with $1\le p<m$, $1\le q<m$ and $1\le p+q\le m$. 
As for the three unitary representations at $c=\frac12, m=3$
(corresponding to $h_{00}, h_{1,2}$ and $h_{21}$), the Verma modules
associated with the others
$h_{pq}$ have singular vectors. The quotient by the subspace
spanned by these vectors is irreducible but the inner
product on the quotient space in not positive definite. Therefore
these representations are not unitary and limits on field indices
appearing in OPE are set to reject them. Prior to the work reported
here, we did not know of any use for these non-unitary representations
in the description of the Ising model, as for example in OPE's.

We describe in each of the following sections an observable for
the Ising model. They are somewhat unconventional as they are
non-local objects. Using the techniques of conformal field theory,
we are able to give predictions for their behavior and calculate
their asymptotic behavior. The latter is described, in the three
cases, by exponents $h_{pq}$ from Kac table that lead to non-unitary
representations.

\goodbreak

\section{Crossing probability\protect\\ on Ising clusters}

The first observable is derived from one in percolation theory and
we start by describing it in this context. In percolation by sites
on a square lattice, each site is declared open (closed) with probability
$p$ (resp.\ $(1-p)$) independently of its neighbors. A configuration on
a finite geometry is said to have a crossing between two disjoint
intervals on the boundary if there is a cluster of open sites joining
the two intervals. A common geometry is the rectangle, say of $m\times
n$ sites, with the two disjoint intervals chosen to be the vertical
sides. One then speaks of a horizontal crossing. A central quantity in
percolation theory is the probability $\pi_h^{\text{perco}}(r)$ of such
crossings when the number of sites goes to infinity, the aspect ratio
$r=\text{width}/\text{height}$ being kept fixed. Cardy \cite{Cardy} gave
a prediction for this function $\pi_h^{\text{perco}}(r)$ using conformal
field theory and the agreement with numerical data is excellent
\cite{lpsa}.

For the Ising model, we define $\pi_h(r)=\pi_h^{\text{Ising}}(r)$ as the
probability of crossing on clusters of $+$ spins, the limit on the
number of sites being taken as above. Lapalme and one the authors have
adapted Cardy's ideas to this case and obtained Monte-Carlo measurements
to test their prediction \cite{Ervig}. Two steps in Cardy's reasoning
cannot be extended straightforwardly to the Ising model and Lapalme and
Saint-Aubin had to resort to one basic property of $\pi_h^{\text{perco}}$
and $\pi_h^{\text{Ising}}$, their scale invariance. If
$\pi_h^{\text{Ising}}$ is described by a four-point correlation function,
it must be that of a field of vanishing conformal weight. They therefore
chose to use the second singular vector of height $6$ in the Verma
module with $c=\frac12, h=0$. The ordinary differential equation
obtained from this singular vector is of order $6$ and the exponents are
$0, \frac16$ (twice degenerate), $\frac12, \frac53$ and $\frac 52$.
If one is willing to ignore the constraints on $p$ and $q$, these
exponents are precisely the conformal weights $h_{pq}$ of the fields
appearing in the (naive) operator product expansion of $\phi_{23}$ of
conformal weight $h_{23}=0$.

\begin{figure}
\begin{center}\leavevmode
\includegraphics[bb = 40 180 380 710,clip,width = 8cm]{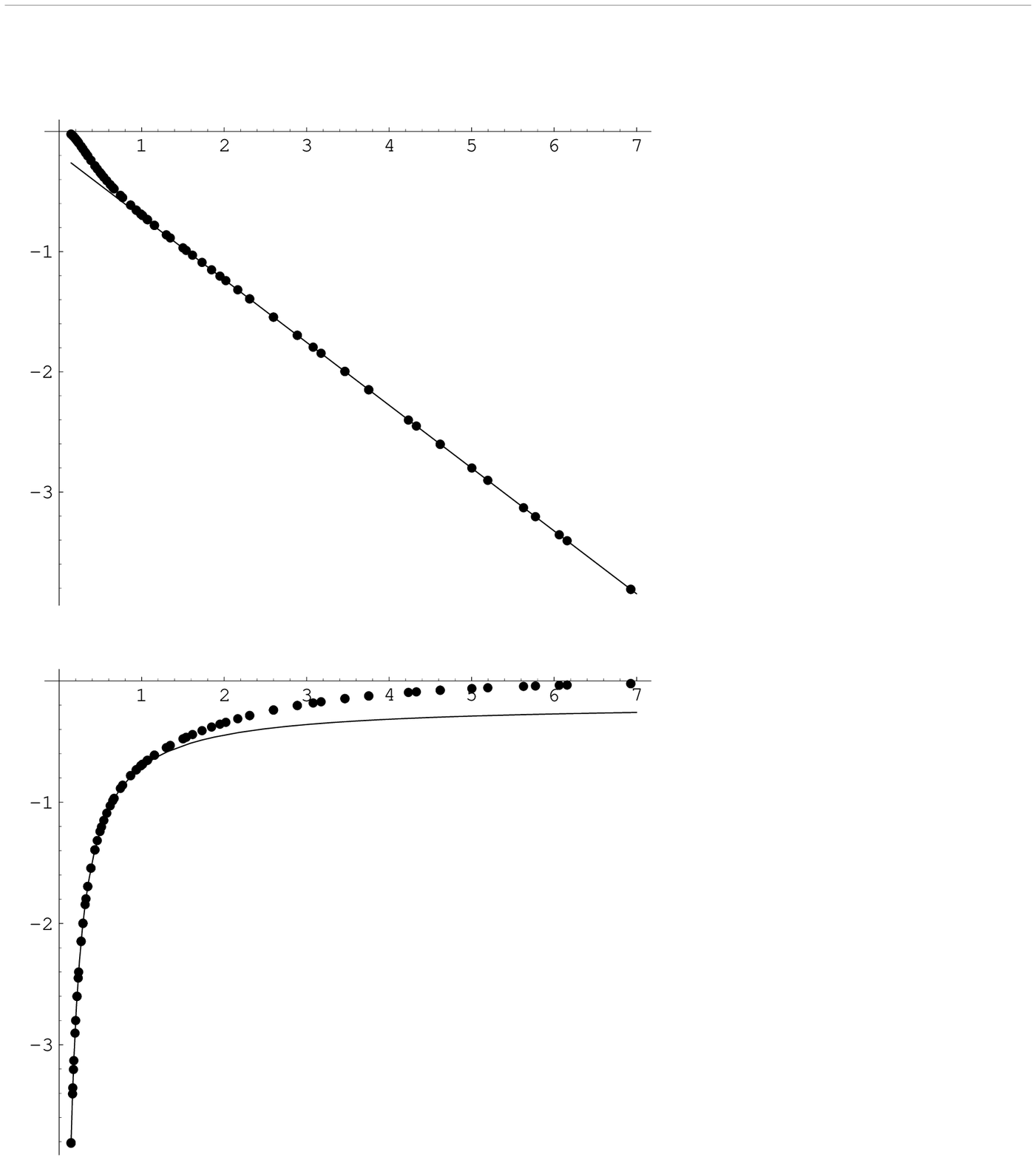}
\end{center}
\caption{Asymptotic behaviors of $\pi_h$ as functions of
$r$. The ordinates are $\log \pi_h(r)$ for the top graph,
$\log(1-\pi_h(r))$ for the bottom one.}\label{fig:pih}
\end{figure}

Lapalme and Saint-Aubin's argument is not as convincing as Cardy's which
relies on better established ideas. The agreement of their prediction with
numerical data is therefore welcome \cite{Ervig}. It is sufficient for the
purpose of this letter to concentrate on the limiting behavior of
$\pi_h(r)$ as
$r\rightarrow0$ and $r\rightarrow\infty$. One expects $\pi_h(r)\rightarrow1$
and $\rightarrow 0$ in these two limits and the leading behavior
prescribed by the ode is that of the exponent $\frac16$. Because
$\frac16$ is twice degenerate, a logarithmic behavior is allowed but appears
to be ruled out by
numerical data. The behaviors that are seen in the simulations are
\begin{align*}
\log \pi_h(r)\rightarrow ar+b, &\qquad r\rightarrow0\\
\log(1-\pi_h(1/r))\rightarrow \frac cr+d,&\qquad r\rightarrow\infty
\end{align*}
with $\hat a=-0.1664\pi$ and $\hat c=-0.1665\pi$. (The factor of $\pi$
stems from the change of variables between the aspect ratio $r$ and the
variable used in the ode. To obtain $\hat a$ and $\hat b$, we used
measurements of both horizontal and vertical crossings in \cite{Ervig}
($\pi_v(r)=1-\pi_h(1/r)$) and fitted the asymptotic behaviors above to the
values of $\pi_h$ for the 20 smallest (largest) aspect ratios $r$. See
Figure \ref{fig:pih}.) 
The measurement errors are of a few units on
the fourth digit. There is little doubt that $\frac16=h_{3,3}$ is the
exponent describing these asymptotics.

\section{Contours intersecting the boundary of a cylinder}

\begin{figure}[b]
  \begin{center}
    \leavevmode
    
    \includegraphics[clip,width = 7.5cm]{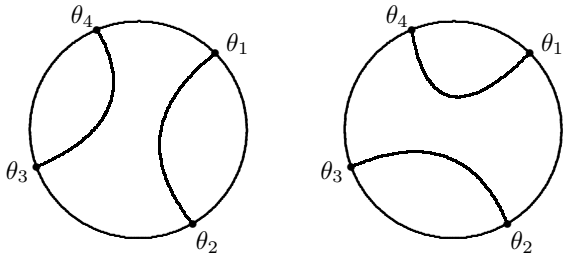}
    \caption{Configurations contributing to different contour probabilities}
    \label{fig:contour}
  \end{center} 
\end{figure}

Consider now the description of the spin $\sigma$ at the boundary of a 
half-infinite cylinder. Let $\theta_1, \theta_2, \dots, \theta_{2n}$ 
be the angles along the boundary where spinflips occur. Using 
Onsager's solution \cite{Onsager} or conformal field theory, it is possible
to calculate the probability density of configurations having precisely
$2n$ flips. For example, this density $s(\theta_1,\theta_2,\theta_3,\theta_4)$
for the case $n=2$ is proportional to
\begin{align*}
(\sin{\textstyle{\frac12}}\theta_{12}\sin{\textstyle{\frac12}}\theta_{34})^{-1}
&-(\sin{\textstyle{\frac12}}\theta_{13}\sin{\textstyle{\frac12}}\theta_{24})^{-1}\\
&+(\sin{\textstyle{\frac12}}\theta_{14}\sin\frac12\theta_{23})^{-1}
\end{align*}
if $\theta_1,\theta_2,\theta_3,\theta_4$ appear in that order along the
boundary and $\theta_{ij}$ is $\theta_i-\theta_j$. This density does not
reveal however which pairs
$(\theta_i,\theta_j)$ are actually joined by the contours between
same-spin clusters. For $n=1$ there is only one possible pairing but,
for $n=2$, there are already $2$. Using conformal invariance, we
represent the half-infinite cylinder by a disk (minus its center) and
depict two possible distinct configurations (see Figure \ref{fig:contour}).
Let $l(\theta_1,\theta_2,\theta_3,\theta_4)$
($r(\theta_1,\theta_2,\theta_3,\theta_4)$) be the density probability for
the pairing of the left (resp.\ right) configuration. Two requirements allow
for the determination of $l$ and $r$ using conformal field theory.
First their sum should reproduce the density when contours are ignored,
namely $l+r=s$. Due to the singularity in $s$, it is natural to seek $l$ 
within the solution space of the ordinary differential equation that 
describes the $4$-point correlation function of the field $\phi_{2,1}$
of conformal weight $\frac12$. Second, when $\theta_{12}\rightarrow
0$, the probability $\text{Prob}({\theta_1\theta_2}|
{\theta_3\theta_4})=l(\theta_1,\theta_2,\theta_3,\theta_4)/
s(\theta_1,\theta_2,\theta_3,\theta_4)$ should go to $1$ and 
$\text{Prob}({\theta_2\theta_3}|{\theta_4\theta_1})=r/s$ to $0$.

These two requirements determine uniquely $l, r$ and 
$\text{Prob}(\theta_1\theta_2|\theta_3\theta_4)$
\cite{Onsager}. For the latter one gets
\begin{align*}
&\text{Prob}({\theta_1\theta_2}|{\theta_3\theta_4})=\\
&\frac12-
\frac9{20}\frac{\Gamma(\frac13)}{\Gamma(\frac23)^2f^{(0)}(z)}\left(
f^{(5/3)}(z)-\frac{z}{1-z}f^{(5/3)}(1-z)\right)
\end{align*}
where the anharmonic ratio is chosen to be
$z={(\sin\frac12\theta_{12}
\sin\frac12\theta_{34})}/{(\sin\frac12\theta_{13}\sin\frac12\theta_{24})}$
and where $f^{(0)}(z)=1-z+z/(1-z)$ and $f^{(5/3)}(z)=z^{\frac53}\ _2F_1(-
{\textstyle{\frac13,\frac43,\frac83}},z)/{(1-z)}$.
The behavior of this prob\-a\-bility as $\theta_{12}\rightarrow 0$ ($z
\rightarrow0$) is 
$$1-\frac{10}9\frac{\Gamma(\frac23)^2}{\Gamma(\frac13)}z^{5/3}+\mathcal{O}(z^2).$$
The exponent $\frac53$ is the highest weight $h_{31}$ of Kac table.

\begin{figure}
\begin{center}\leavevmode
\includegraphics[bb = 55 425 535 725,clip,width = 8cm]{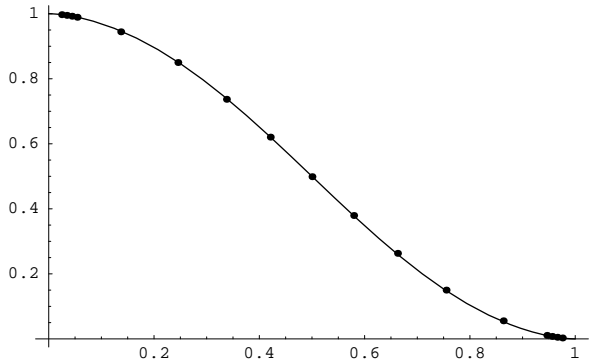}
\end{center}
\caption{The function $\text{Prob}({\theta_1\theta_2}|
{\theta_3\theta_4})$ as function of $z\in[0,1]$ with Monte-Carlo
data.}\label{fig:prob}
\end{figure}

The function
$\text{Prob}({\theta_1\theta_2}|{\theta_3\theta_4})$ is plotted
on Figure \ref{fig:prob} together with Monte-Carlo measurements of this
probability on a cylinder whose length is twice as long as its
circumference. The four dots close to $z=0$ and the four close to $z=1$
were measured on a cylinder with $32000$ sites and samples were larger
than $10^6$ configurations. Statistical errors on these $8$ points are
smaller than the size of the dots on the figure. (Details will be given
in \cite{Onsager}.)

\section{Homology class\protect\\ of Fortuin-Kasteleyn clusters}

\begin{figure}[htbp]
  \begin{center}
    \leavevmode
    
    \includegraphics[clip,width = 8cm]{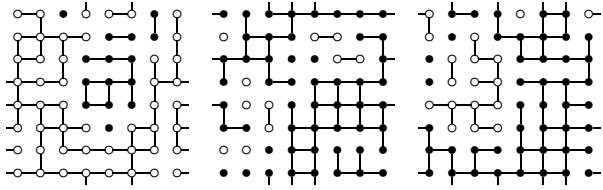}
    \caption{Fortuin-Kasteleyn clusters for three
        configurations. They contribute to 
    $\pi\{(1,0)\}$, $\pi\{(1,-1)\}$ and $\pi(\text{cross})$ respectively.}
    \label{fig:graphes}
  \end{center} 
\end{figure}

The last observable describes the Ising model on a geometry without
boundary. Denote by $\alpha$ ($\beta$) a non-trivial cycle in the
horizontal (vertical) direction on a torus of modulus $\tau$. Let
$a,b\in\mathbb{Z}$ be two integers with $\gcd(a,b)=1$ and let
$\pi(\{a,b\})$ be the probability that a spin configuration has
a Fortuin-Kasteleyn cluster wrapping precisely $a$ times around $\alpha$
and $b$ around $\beta$. (Configurations having simultaneously clusters of
type
$\{1,0\}$ and $\{0,1\}$ are not included in the computation of
$\pi(\{a,b\})$. These configurations are said to contain a cross.) 
In Figure \ref{fig:graphes} the Fortuin-Kasteleyn clusters of
three configurations on a square torus ($\tau = i$)
are shown. The original signs of the Ising spins
are depicted in black ($-$) or in white ($+$).
Only the leftmost configuration in this Figure contributes to $\pi(\{1,0\})$.  

Using the Coulomb gas representation \cite{diFSZ, Pinson} it
is possible to write an explicit formula for $\pi(\{a,b\})$
(see \cite{Arguin} where this expression is given for
various Potts models and where Monte-Carlo simulations
checking it are reported). In particular for the Ising model
\begin{equation*}\label{eq:pi10}
\pi_\tau(\{(1,0\})=\frac1{|\eta(q^2)|}\frac{\theta_2(\frac i3\tau_i)
-|\theta_3(\frac i3\tau_i)-\theta_4(\frac i3\tau_i)|}{
{|\theta_2(\tau)|+|\theta_3(\tau)}|+|\theta_4(\tau)|}
\end{equation*}
where $\tau_i=\text{Im}\, \tau$, $q=e^{i\pi\tau}$, the $\theta_i$ are
the elliptic theta functions and $\eta(q)$ is Dedekind function
$\eta(q)=q^{\frac1{24}}\prod_{n=1}^\infty (1-q^n)$. For $\text{Re}\,\tau
=0$ and $\tau_i\rightarrow\infty$, that is for a torus represented by 
a very narrow and tall rectangle, any configuration will almost
surely contain a horizontal cluster and no vertical one. In that
limit one expects that all $\pi_\tau(\{(a,b\})$ will vanish, including
$\pi_\tau(\text{cross})$, except for $\pi_\tau(\{1,0\})$. For that case
$q=e^{-\pi\tau_i}$ and one gets indeed as $\tau_i\rightarrow\infty$
\begin{equation*}
\pi_\tau(\{1,0\})\rightarrow 1-(q^2)^{\frac18}f_1(q^2)-
(q^2)^{\frac13}f_2(q^2)-\dots
\end{equation*}
where the $f_i$ are real analytic in their argument in a neighborhood
of $0$. The two leading terms
are recognized to be twice the weights $h_{1,2}$ and $h_{3,3}$.
(The doubling of highest weights in expansions in $q^2$ 
is the natural thing to expect and
accounts for the contributions of holomorphic and antiholomorphic
sectors of the theory.) 

Of the three 
observables discussed in this note, this one ($\pi_\tau(\{1,0\})$) is probably
the less compelling. There is no differential equation here to
dictate a {\em finite} set of exponents. The ``$\dots$'' in 
the above asymptotic expansion contains other exponents, namely
other integral linear combinations of $\frac18$ and $\frac13$.
Not all these combinations however occur in Kac table.  (We showed
that the exponents are restricted to the
set $\{\frac n8, n=0,1,\dots, 7\}\cup\{\frac13+\frac n8,n=0,1,\dots, 7\}$
but we did not prove that all of these do occur.)

Despite these
comments, the fact that the highest weights $h_{12}$ and $h_{33}$
describe the leading behavior seems remarkable.

\section{Conclusion}

What are the possible
exponents of the Kac table at $m=3$ if one
identifies
$h_{pq}$ and $h_{p'q'}$ whenever $h_{pq}-h_{p'q'}\in\mathbb{Z}$?
This amounts to asking the simple number theoretic question of which
integers modulo $48$ have a square root. The answer to the first question
is the following list of $h$'s: the three unitary $0,\frac1{16}$ and
$\frac12$ and the non-unitary $\frac53,\frac16, \frac5{16},\frac{35}{48}$
and $-\frac1{48}$. The question of whether other observables can be
constructed that are ruled by $\frac5{16},\frac{35}{16}$ and
$-\frac{1}{48}$ is more difficult.

More to the point are the following questions. Which linear space of
states must be considered so that non-local observables may be taken into
account in the framework of conformal field theory? Why do statistical
models seem to prefer unitary representations even when mutually non-local
fields are considered (like the pair $\sigma$--$\mu$ in the Ising
model)? And why, when they do step out of the unitary representations,
do they remain in Kac table?

\bigskip

\hrule

\bigskip

\centerline{\bfseries NOTE}

After the research reported here was completed, one of us (YSA) learned from
Duplantier that various exponents for the Ising model
have been introduced by him and his colleagues
that do not belong to the small set $\{0, \frac1{16},\frac12\}$.

Here are two representative examples discussed in their work. For $O(n)$
models, consider the probability density of having $L$ loops between
two points $x$ and $y$ in the plane. Duplantier \cite{Du1} shows that
it decays as $|x-y|^{-2x_L}$ with $x_L=2h_{L/2,0}$. Not only do these
exponents miss the unitary set $\{0, \frac1{16},\frac12\}$ but, for $L$
odd, they are out of Kac table. This fact raises questions beyond the
present letter.

More recently \cite{Du2}, Duplantier obtained the fractal dimensions
of three properties of Potts clusters, namely of the external perimeter
($D_{EP}$), of the set of outer boundary sites (dimension $D_H$ of the
hull) and of the singly connecting sites that appear close under the
scaling limit ($D_{SC}$). For the Ising model he gets $D_{EP}=\frac{11}8$,
$D_H=\frac53$ and $D_{SC}=\frac{13}{24}$. It would be hard to believe
that the $\frac 53$ is a coincidence but the other two dimensions are out
of Kac table. The role of these exponents for the argument presented here
is unclear to us.

The most intriguing connection with the present work has appeared
recently. Read and Saleur \cite{ReadSa} consider nonlinear sigma models 
whose fields take values in supersymmetric coset spaces. They argue that, 
for the target space $S^{2n|2n}$ (a supersymmetric generalization of the
sphere), the model shows a Ising-like transition. The spectrum of its conformal
weights is described by their formula (3.7--8) with $e_0f=\frac16$ and
$gf^2=\frac13$ and includes the set $\{0,\frac1{16},\frac12\}$ and the
exponents observed here: $\frac53$ and $\frac16$. The relationship
between their models and the observables discussed here remains to be
established.

\bigskip

\hrule

\bigskip 

\section*{Acknowlegments}
The authors would like to thank R.P.\ Langlands, F.\ Lesage and P.\ Mathieu
for useful discussions.
 
L.-P.\ A.\ gratefully acknowledges a fellowship from the NSERC Canada  
Scholarships Program and Y.\ S.-A.\ support 
from NSERC (Canada). Y.\ S.-A.\ also thanks the Institute for Advanced
Study (Princeton) for its hospitality and generosity.

\end{document}